\documentclass[prl,floatfix,showpacs,twocolumn,preprintnumbers,amsmath,amssymb,superscriptaddress]{revtex4}
\usepackage{graphicx,float,datetime}
\usepackage[ansinew]{inputenc}

\newcommand{\dut}[3]{#1_{#2}^{\phantom{#2}#3}}

\newcommand{\Lagr}{\mathcal{L}}

\newdateformat{mydate}{\THEDAY\hspace{3pt}\monthname[\THEMONTH] \THEYEAR}

\begin{document}

\begin{center}
\title{Gravitomagnetic effects in conformal gravity}
\date{\mydate\today}
\author{Jackson Levi Said\footnote{jsai0004@um.edu.mt}}
\affiliation{Physics Department, University of Malta, Msida, MSD 2080, Malta}
\author{Joseph Sultana\footnote{joseph.sultana@um.edu.mt}}
\affiliation{Mathematics Department, University of Malta, Msida, MSD
2080, Malta}
\author{Kristian Zarb Adami\footnote{kristian.zarb-adami@um.edu.mt}}
\affiliation{Physics Department, University of Malta, Msida, MSD 2080, Malta}
\affiliation{Physics Department, University of Oxford, Oxford, OX1 3RH, United Kingdom}

\begin{abstract}
{Gravitomagnetic effects are characterized by two phenomena: first, the geodetic effect which describes the precession of the spin of a gyroscope in a free orbit around a massive object, second, the Lense-Thirring effect which describes the precession of the orbital plane about a rotating source mass. We calculate both these effects in the fourth-order theory of conformal Weyl gravity for the test case of circular orbits. We show that for the geodetic effect a linear term arises which may be interesting for high radial orbits, whereas for the Lense-Thirring effect the additional term has a diminishing effect for most orbits. Circular orbits are also considered in general leading up to a generalization of Kepler's third law.}
\end{abstract}


\maketitle

\end{center}

\section{I. Introduction}
The validity of any alternative theory to Einstein's general
relativity depends on how well it agrees with his theory in the weak field limit as well as observational tests. One of the possible alternatives to Einstein's second order theory which has been proposed during the last two decades is conformal Weyl gravity \cite{conformal1,
conformal2,conformal3}. Instead of choosing the gravitational action by requiring that the theory be no higher than second order as in
the case of the Einstein-Hilbert action, Weyl gravity employs the principle of local conformal invariance of spacetime to fix the gravitational action, meaning that the theory is invariant under local conformal stretching of the type
\begin{equation}
g_{\mu\nu}(x) \rightarrow \Omega^2(x) g_{\mu\nu}(x),
\end{equation}
where $\Omega(x)$ is a smooth strictly positive function. This leads to a conformally invariant fourth order theory with a unique action given by
\begin{eqnarray} I_{W} & = & -\alpha\int
d^4x(-g)^{1/2}C_{\lambda\mu\nu\kappa}C^{\lambda\mu\nu\kappa} \nonumber\\
      & = & -2\alpha\int d^4x(-g)^{1/2}[R_{\mu\kappa}R^{\mu\kappa} -
      (R^{\nu}_{\nu})^2/3] \nonumber\\
      & & +{\rm a~total~derivative}, \label{action}
\end{eqnarray}
where $C_{\lambda\mu\nu\kappa}$ is the conformal Weyl tensor and
$\alpha$ is a purely dimensionless coefficient. The corresponding
field equations are then given by \cite{conformal1}
\begin{eqnarray}
\sqrt{-g}g_{\mu\alpha}g_{\nu\beta}\frac{\delta I_{W}}{\delta
g_{\alpha\beta}}
&=-2\alpha W_{\mu\nu}\nonumber\\
&=-\frac{1}{2}T_{\mu\nu}, \label{weyl_field_eqns}
\end{eqnarray}
where $T_{\mu\nu}$ is the stress-energy tensor, and
\begin{equation}
W_{\mu\nu} = 2C^{\alpha\ \ \beta}_{\ \mu\nu\ ;\beta\alpha} +
C^{\alpha\ \ \beta}_{\ \mu\nu}R_{\alpha\beta}. \label{w1}
\end{equation}
It can be seen from (\ref{weyl_field_eqns}) that
any vacuum solution of Einstein's field equations is also a vacuum
solution of Weyl gravity; with the converse not necessarily being true.
Despite being a fourth-order theory with highly nonlinear field
equations, a number of exact solutions
\cite{furthersolutions1,furthersolutions2,furthersolutions3,
furthersolutions4,cylindrical1,cylindrical121} have been found which
generalize the well-known Kerr-Newmann and cylindrical solutions of
general relativity.
\newline
The exact static and spherically symmetric vacuum solution for
conformal gravity is given, up to a conformal factor, by the metric
\cite{conformal1}
\begin{equation}
ds^2 = -B(r)dt^2 + \frac{dr^2}{B(r)} + r^2(d\theta^2 + \sin^2\theta
d\phi^2), \label{generalmetric}
\end{equation}
where
\begin{equation}
B(r) = 1 - \frac{\beta(2 - 3\beta\gamma)}{r} - 3\beta\gamma + \gamma
r - k r^2, \label{eq:metric}
\end{equation}
and $\beta,\ \gamma,\  \mbox{and}\ k$ are integration constants.
This solution includes as special cases the Schwarzschild solution
$(\gamma = 0 = k)$ and the Schwarzschild-de Sitter $(\gamma = 0)$
solution; the latter requiring the presence of a cosmological
constant in Einstein gravity. Moreover the constant $\gamma$ has
dimensions of acceleration, and so the solution provides a
characteristic, constant acceleration without having introduced one
at the Lagrangian (such as in the relativistic implementation of
MOND with TeVeS \cite{bekenstein}).
\newline
The magnitude and the origin (i.e. whether it is system dependent
like $\beta$ or cosmological like $k$) of the integration constant
$\gamma$ in (\ref{eq:metric}) remains unknown. When it is associated
with the inverse Hubble length, i.e. $\gamma \simeq 1/R_H$, the
effects of the acceleration  due to the $\gamma r$ term in the
metric are comparable with those due to the Newtonian potential term
$2\beta/r \equiv r_s/r$ ($r_s$ is the Schwarzschild radius), on
length scales given by
\begin{equation}
r_s / r^2 \simeq \gamma \simeq 1/R_H ~~{\rm or}~~ r \simeq (r_s \,
R_H)^{1/2}. \label{eq:MRrelation}
\end{equation}
As noted in \cite{conformal1}, for a galaxy of mass $M \simeq
10^{11}\; {\rm M}_{\odot}$ with  $r_s \simeq 10^{16}$ cm and $R_H
\simeq 10^{28}$ cm, this scale is $r \sim 10^{22}$ cm, i.e. roughly
the size of the galaxy, a fact that prompted Mannheim, O'Brien and Kazanas to
produce fits to the galactic rotation curves using the metric of Eq.
(\ref{eq:metric}) above, without the need to invoke the presence of
dark matter as in standard Einstein's theory (see Refs.
\cite{mannheim11,obrien12} for recent work on this issue).
\newline
However, an issue arises in that in addition to having flat rotation curves in the region of interest for galaxies, we must also have stability for the orbits under consideration \cite{Nandi}. Circular orbits offer a good toy model for investigating different gravitational effects in alternative theories of gravity. Due to their importance we begin with a short investigation of circular orbits leading up to a statement of Kepler's third law.
\newline
The geodetic precession, also known as de Sitter precession is a general relativistic phenomenon discovered in 1916 by de Sitter \cite{desitter16}, who found that the spin of a gyroscope precesses with respect to a distant inertial frame when it makes free (or ``geodetic'') orbits around a nonrotating massive object. For a circular orbit the amount of geodetic precession per orbit is given by
\begin{eqnarray}
\Delta\phi_{\mbox{geodetic}} & = & 2\pi\left[1 - \left(1 -
\frac{3GM}{c^2R}\right)^{1/2}\right] \nonumber \\
& \approx &  \frac{3\pi GM}{c^2R}, \quad \mbox{for}\ \frac{GM}{c^2
R} << 1,
\end{eqnarray}
where $M$ is the mass of the massive object and $R$ is the radius of
the orbit. This effect is also observed in flat spacetime, where in
this case it has a purely kinematical origin and is known as Thomas
precession \cite{thomas26}. Another closely related relativistic
precession is the Lense-Thirring effect \cite{lense18} (or frame
dragging effect) discovered by Lense and Thirring in 1918, and
refers to the precession of the gyroscope due to the rotation of the
massive object which produces a dragging of nearby inertial frames.
Although these are two independent effects in the sense that the
latter requires rotation of the source of the gravitational field,
it can be shown that in an appropriately chosen coordinate system
\cite{ashby90}, geodetic precession can be considered as due to a
Lense-Thirring drag. A combined observable phenomenon of these two
effects \cite{schiff60} occurs in the Earth-Moon system around the
Sun through the precession of the Moon's perigee, which is detected
by measuring the lunar orbit using laser ranging between stations on
Earth and reflectors on the Moon's surface
\cite{bertotti87,shapiro88,dickey94}.
\newline
A space experiment to test these two predictions of Einstein's theory is Gravity Probe B (GP-B) \cite{everitt11} which was launched on 20 April 2004 in a 642 km polar Earth orbit with data collection lasting almost a year. Measurement of the geodetic and frame-dragging effect was done by means of cryogenic gyroscopes with one or more of these referenced to a remote star by means of an onboard telescope. For the chosen orbit the two effects result in a precession along two perpendicular planes so that GP-B could measure these independently. Analysis of the data from the four onboard gyroscopes showed a geodetic drift rate of $-6601.8 \pm 18.3\ \mbox{mas/yr}$ and a frame dragging drift rate of $-37.2 \pm 7.2\ \mbox{mas/yr}$, which are in accordance with the general relativistic predictions of $-6606.1\ \mbox{mas/yr}$ and $-39.2\ \mbox{mas/yr}$ respectively.
\newline
The classical tests of general relativity including the bending of light \cite{pireaux04a,pireaux04b,amore06,sultana10}, time delay \cite{edery98} and perihelion precession \cite{sultana12} have already been studied in conformal gravity. In this paper we investigate circular orbits in Sec. II, deriving Kepler's third law. In Sec. III the geodetic precession effect is determined for circular orbits. In Sec. IV the Lense-Thirring effect is calculated in terms of the precession velocity. Finally in Sec. V we end with a summary of conclusions and a discussion.

\section{II. Circular orbits and Kepler's Third Law}
We investigate circular orbits in conformal gravity using a Lagrangian approach for the metric in Eq.(\ref{generalmetric}). In this case the Lagrangian takes the form
\begin{align}
\Lagr=&\frac{1}{2}g_{\mu\nu}\frac{\partial x^{\mu}}{\partial\tau}\frac{\partial x^{\nu}}{\partial\tau}\nonumber\\
=&\frac{1}{2}\left[-B\dot{t}^2+\frac{\dot{r}^2}{B}+r^2\dot{\theta}^2+r^2\sin^2\theta\dot{\phi}^2\right],
\label{613_lagrangian}
\end{align}
where dots denote differentiation with respect to proper time, $\tau$. Using the four-velocity normalization condition $u_{\mu}u^{\mu}=-1$, where
$u_{\mu}\equiv\frac{dx^{\mu}}{d\tau}$, it follows that $\Lagr=-\frac{1}{2}$. It must be noted at this point that this Lagrangian and the
associated timelike geodesics representing trajectories of free massive particles are not conformally invariant, unlike the theory of conformal
gravity. Moreover conformal invariance implies that there is no prescribed scale in the theory, which has a dimensionless coupling constant $\alpha$ as shown in Eq. (\ref{action}). However recently Edery et al. in Ref. \cite{edery06} (see also Refs. \cite{bouchami08,matsuo90}) considered spontaneous symmetry
breaking in Weyl gravity and showed that this can provide a mechanism to generate a scale in the theory.
\newline
Taking advantage of the independence of the Lagrangian on the $t-$ and $\phi-$coordinates, and using $\theta = \pi/2$ as the plane of the circular orbits without loss of generality due to spherical symmetry, the Euler-Lagrange equations of motion are given by
\begin{align}
B\dot{t}&=E,\\
r^2\dot{\phi}&=L,
\end{align}
 where $E$ and $L$ are the energy and angular momentum respectively.
The third Euler-Lagrange equation is given by
\begin{align}
& &(B^{-1}2\dot{r}\dot{)} - [-(\frac{\beta(2 - 3\beta\gamma)}{r^2} + \gamma - 2kr)\dot{t}^2  \nonumber\\
& &+\frac{\partial}{\partial r}(B^{-1})\dot{r}^2 + 2r\dot{\phi}^2] = 0.
\end{align}
For circular orbits $r = R$, $\dot{r} \equiv 0$ and therefore the above equation yields
\begin{equation}
\omega^2 = \left(\frac{d\phi}{d t}\right)^2 = \frac{1}{2R}\left(\frac{\beta(2 - 3\beta\gamma)}{R^2} + \gamma - 2kR\right).
\end{equation}
For $k << \gamma << \beta$ this can be written as
\begin{equation}
\omega^2 \sim \frac{\beta}{R^3} + \frac{\gamma}{2R},
\end{equation}
which is similar to Kepler's third law, with the second term on the right-hand-side representing the correction from conformal gravity.

\section{III. The Geodetic effect}
The precession is calculated for orbits in the equatorial plane
$\theta=\frac{\pi}{2}$ since by symmetry considerations a rotation
can always be made to this plane. Rotating coordinates are first
introduced through the transformation
\begin{equation}
\phi\rightarrow\phi-\omega t, \label{trans}
\end{equation}
where $\omega$ is the coordinate angular frequency of the rotation.
This transforms the metric in Eq.(\ref{eq:metric}) to
\begin{widetext}
\begin{align}
ds^2&=&-\left(1-\frac{\beta\left(2-3\beta\gamma\right)}{r}-3\beta\gamma+\gamma r-kr^2-r^2\omega^2\right)\left[dt-\frac{r^2\omega}{1-\frac{\beta\left(2-3\beta\gamma\right)}{r}-3\beta\gamma+\gamma r-kr^2-r^2\omega^2}d\phi\right]^2\nonumber\\
& &
+\frac{dr^2}{1-\frac{\beta\left(2-3\beta\gamma\right)}{r}-3\beta\gamma+\gamma
r-kr^2}+r^2\frac{1-\frac{\beta\left(2-3\beta\gamma\right)}{r}-3\beta\gamma+\gamma
r-kr^2}{1-\frac{\beta\left(2-3\beta\gamma\right)}{r}-3\beta\gamma+\gamma
r-kr^2-r^2\omega^2}d\phi^2.
\end{align}
\end{widetext}
Comparing this with Rindler's canonical form \cite{b04}
\begin{equation}
ds^2=-e^{2\Phi}\left(dt-w_idx^i\right)^2+k_{ij}dx^i dx^j,
\label{canonical}
\end{equation}
the following nonvanishing components arise
\begin{equation}
e^{2\Phi}=1-\frac{\beta\left(2-3\beta\gamma\right)}{r}-3\beta\gamma+\gamma r-kr^2-r^2\omega^2 \label{eqn1}
\end{equation}
\begin{equation}
w_3=\frac{r^2\omega}{1-\frac{\beta\left(2-3\beta\gamma\right)}{r}-3\beta\gamma+\gamma r-kr^2-r^2\omega^2} \label{eqn2}
\end{equation}
\begin{equation}
k^{11}=1-\frac{\beta\left(2-3\beta\gamma\right)}{r}-3\beta\gamma+\gamma r-kr^2
\end{equation}
\begin{equation}
k^{33}=\frac{1-\frac{\beta\left(2-3\beta\gamma\right)}{r}-3\beta\gamma+\gamma r-r^2\left(k+\omega^2\right)}{r^2\left(1-3\beta\gamma\right)-\beta r\left(2-3\beta\gamma\right)+r^3\gamma-kr^4}, \label{eqn3}
\end{equation}
where Latin indices refer to spacelike coordinates only.
\newline
We consider free circular orbits for which the acceleration
\cite{b01} vanishes
\begin{equation}
a=\left(k^{ij}\Phi_{,i}\Phi_{,j}\right)^{1/2}=0,
\end{equation}
which implies that $\Phi_{,r}=0$. Hence a relation can be
established between the angular velocity and the metric parameters
such that
\begin{equation}
\omega^2=\frac{\beta\left(2-3\beta\gamma\right)}{2r^3}+\frac{\gamma}{2r}-k.
\end{equation}

Substituting this in Eqs.(\ref{eqn1}, \ref{eqn2}, \ref{eqn3}) gives
\begin{align}
e^{2\Phi}&=1-\frac{3\beta\left(2-3\beta\gamma\right)}{2r}-3\beta\gamma+\frac{\gamma r}{2} \label{e2phi}\\
w_3&=\frac{r^2\sqrt{\frac{\beta\left(2-3\beta\gamma\right)}{2r^3}+\frac{\gamma}{2r}-k}}{1-\frac{3\beta\left(2-3\beta\gamma\right)}{2r}-3\beta\gamma+\frac{\gamma r}{2}} \label{w3}\\
k^{33}&=\frac{1-3\beta\,\frac{2-3\beta\gamma}{2r}-3\beta\gamma+\frac{r\gamma}{2}}{r^2\left(1-3\beta\gamma\right)-\beta
r\left(2-3\beta\gamma\right)+r^3\gamma-kr^4}. \label{k33}
\end{align}
\newline
The proper rotation rate of a gyrocompass with respect to the
rotating frame (\ref{trans}) is given in terms of the canonical form
(\ref{canonical}) by \cite{b01}
\begin{align}
\Omega&=\frac{1}{2\sqrt{2}}e^{\Phi}\left[k^{ik}k^{jl}\left(w_{i,j}-w_{j,i}\right)\left(w_{k,l}-w_{l,k}\right)\right]^{1/2}\nonumber\\
&=\frac{e^{\Phi}}{2}\left[k^{11}k^{33}\dut{w}{3,1}{2}\right]^{1/2}.
\end{align}
Substituting Eqs. (\ref{e2phi} - \ref{k33}) in this expression we
get $\Omega = \omega$,  so that the coordinate rate $\omega$ is
really the orbital rotational frequency.
\newline
Now for the Mannheim-Kazanas metric (\ref{eq:metric}) the relation
between proper and coordinate times for a circular orbit
($r = R,\ \dot{r}=0$) with coordinate angular velocity $\omega$, is given by
\begin{equation}
\Delta\tau=\sqrt{1-\frac{3\beta\left(2-3\beta\gamma\right)}{2R}-3\beta\gamma+\frac{\gamma
R}{2}-2R^2k}\;\;\Delta t.
\end{equation}

Hence for one complete orbit about the gravitating mass the
precession of the gyroscope with respect to the rotating frame is
\begin{align}
\alpha'&=\Omega\Delta\tau\nonumber\\
&=2\pi\sqrt{1-\frac{3\beta\left(2-3\beta\gamma\right)}{2R}-3\beta\gamma+\frac{\gamma R}{2}-2R^2k}.
\end{align}

This results in a precession angle per orbit of
\begin{align}
\alpha&=2\pi-\alpha'\nonumber\\
&\approx2\pi\left(\frac{3\beta\left(2-3\beta\gamma\right)}{4R}+\frac{3\beta\gamma}{2}-\frac{\gamma
R}{4}+R^2k\right), \label{precession}
\end{align}
with respect to the inertial frame, where the relation
$u^{\phi}=\omega=\frac{2\pi}{\Delta t}$ was used.

\section{IV. Lense-Thirring Effect}
The Lense-Thirring effect describes the correction of a rotating spacetime on the precession of orbits. The effect can be quantified through a consideration of the Sagnac effect as in \cite{Tartaglia,Nandi,Ruggiero}, which represents the difference in travel time or phase shift of corotating and counterrotating light waves in the field of a central massive and spinning object.  In order to calculate this in conformal gravity we must consider the rotating metric found in \cite{furthersolutions1}, namely
\begin{align}
ds^2&=\left(bf-ce\right)\left(\frac{dr^2}{a}+\frac{dy^2}{d}\right)\nonumber\\
&+\frac{1}{bf-ce}\left[d\left(b\,d\phi-c\,dt\right)^2-a\left(e\,d\phi-f\,dt\right)^2\right],
\end{align}
which is the canonical Carter form of the metric, where
\begin{align}
a&=j^2+ur+pr^2+vr^3-kr^4,\\
d&=1+r'y-py^2+sy^3-j^2ky^4,\\
b&=j^2+r^2,\\
e&=j\left(1-y^2\right),\\
c&=j,\\
f&=1,
\end{align}
and $u,\ p,\ v,\ k,\ r'$ and $s$ are constants satisfying the constraint $uv-r's=0$. The angular momentum of the source is represented by $j$. The general relativistic result representing the Kerr solution is recovered when
\begin{align}
u&=-2MG/c^2,\nonumber \\
p&=1, \nonumber\\
v&=0=k=r'=s,
\label{gr_limit}
\end{align}
while the Kerr-de Sitter solution is obtained when $k  = \Lambda/3$ and $p = 1 - kj^2$; $\Lambda$ being the cosmological constant and with the other parameters taking the same values as in (\ref{gr_limit}).
For the conformal case the additional parameters must be constrained through observation.

Similar to the general relativistic case we take the transformation
\begin{equation}
y\rightarrow\cos\theta,
\end{equation}
in order to obtain Boyer-Lindquist-like coordinates. Noting that $dy=-\sin\theta d\theta$ the metric turns out to be
\begin{widetext}
\begin{align}
ds^2&=\left(r^2+j^2\cos^2\theta\right) \left(\frac{dr^2}{j^2+ur+pr^2+vr^3-kr^4}+\frac{\sin^2\theta\,d\theta^2}{1+r'\cos\theta-p\cos^2\theta+s\,\cos^3\theta-j^2k\cos^4\theta}\right)\nonumber\\
&+\frac{1}{r^2+j^2\cos^2\theta}\Big[\left(1+r'\cos\theta-p\cos^2\theta+s\cos^3\theta-j^2k\cos^4\theta\right)\left(\left(j^2+r^2\right)\,d\phi-j\,dt\right)^2\nonumber\\
&-\left(j^2+ur+pr^2+vr^3-kr^4\right)\left(j\sin^2\theta\,d\phi-dt\right)^2\Big]. \label{kerr-like}
\end{align}
\end{widetext}

The Sagnac effect is now investigated by considering corotating and counterrotating circular light beams in the equatorial plane $\theta = \frac{\pi}{2}$ about the central object. After a circular orbit these light beams reach a detector (that can also double as a source) which is assumed to be rotating with uniform angular velocity $\omega_0$, such that its rotation angle is
\begin{equation}
\phi_0=\omega_0 t.
\label{detector eqn}
\end{equation}
Since we are considering circular orbits, we also set $r=R$ so that
\begin{align}
ds^2&=\frac{dt^2}{R^2}\Big[\left(\left(j^2+R^2\right)\omega_0-j\right)^2\nonumber\\
&-\left(j^2+uR+pR^2+vR^3-kR^4\right)\left(j\omega_0-1\right)^2\Big].
\label{light_metric_rot}
\end{align}

Now given that we are dealing with rays of light we take $ds=0$. Eq.(\ref{light_metric_rot}) then gives two solutions for $\omega_0$, where
\begin{widetext}
\begin{align}
&\Omega_{\pm}=\frac{1}{2\left(R^4+j^2\left(-uR+\left(2-p\right)R^2-vR^3+kR^4\right)\right)}\Big[-2j\left(uR+\left(p-1\right)R^2+vR^3-kR^4\right)\nonumber\\
&\pm\sqrt{4j^2\left(uR+\left(p-1\right)R^2+vR^3-kR^4\right)^2+4\left(R^4+j^2\left(-uR+\left(2-p\right)R^2-vR^3+kR^4\right)\right)\left(uR+pR^2+vR^3-kR^4\right)}\Big],
\end{align}
\end{widetext}
these two solutions refer to the rotating and counterrotating orbits.

For light the rotation angles are given by
\begin{equation}
\phi_{\pm}=\Omega_{\pm}t.
\end{equation}

In conjunction with Eq.(\ref{detector eqn}) the $t-$coordinate can be eliminated
\begin{equation}
\phi_{\pm}=\frac{\Omega_{\pm}}{\omega_0}\phi_0.
\label{angular_vel_with_pi}
\end{equation}

The first intersection of the rays of light with the position of the observer after emission at $t=0$, occurs when
\begin{align}
\phi_+&=\phi_0+2\pi,\\
\phi_-&=\phi_0-2\pi,
\end{align}
which when substituted back into Eq.(\ref{angular_vel_with_pi}) gives
\begin{equation}
\frac{\Omega_{\pm}}{\omega_0}\phi_0=\phi_0\pm2\pi.
\end{equation}
Solving for $\phi_0$
\begin{equation}
\phi_{0_{\pm}}=\pm\frac{2\pi\omega_0}{\Omega_{\pm}-\omega_0}=\pm2\pi\omega_0\,\xi_{\pm},
\end{equation}
where
\begin{widetext}
\begin{align}
&\xi_{\pm}=\left[2\left(R^4+j^2\left(-uR+\left(2-p\right)R^2-vR^3+kR^4\right)\right)\right]\Big[-2j\left(uR+\left(p-1\right)R^2+vR^3-kR^4\right)\nonumber\\
&\pm\sqrt{4j^2\left(uR+\left(p-1\right)R^2+vR^3-kR^4\right)^2+4\left(R^4+j^2\left(-uR+\left(2-p\right)R^2-vR^3+kR^4\right)\right)\left(uR+pR^2+vR^3-kR^4\right)}\nonumber\\
&-2\omega_0\left(R^4+j^2\left(-uR+\left(2-p\right)R^2-vR^3+kR^4\right)\right)\Big]^{-1}.
\end{align}
\end{widetext}

Combining this with the metric in Eq.(\ref{light_metric_rot}), and reverting to physical units the proper time delay can thus be determined by using the detector coordinate relation in Eq.(\ref{detector eqn})
\begin{widetext}
\begin{equation}
d\tau= \frac{d\phi}{c R\omega_0}\sqrt{\left(j^2+uR+pR^2+vR^3-kR^4\right)\left(j\omega_0-1\right)^2-\left(\left(j^2+R^2\right)\omega_0-j\right)^2}
\end{equation}
\end{widetext}

Integrating between $\phi_{0_-}$ and $\phi_{0_+}$ gives the Sagnac time delay between the two light beams
\begin{widetext}
\begin{align}
\delta\tau&=\frac{\phi_{0_+}-\phi_{0_-}}{cR\omega_0}\sqrt{\left(j^2+uR+pR^2+vR^3-kR^4\right)\left(j\omega_0-1\right)^2-\left(\left(j^2+R^2\right)\omega_0-j\right)^2}\nonumber\\
&=\frac{2\pi\left(\xi_++\xi_-\right)}{cR}\sqrt{\left(j^2+uR+pR^2+vR^3-kR^4\right)\left(j\omega_0-1\right)^2-\left(\left(j^2+R^2\right)\omega_0-j\right)^2}.
\end{align}
\end{widetext}
In order to calculate the correction due solely to the rotation of the source the detector rotational parameter, $\omega_0$, is set to zero, such that
\begin{equation}
\delta\tau_0=\frac{4\pi j}{cR}\frac{uR+\left(p-1\right)R^2+vR^3-kR^4}{\sqrt{uR+pR^2+vR^3-kR^4}},
\end{equation}
which can be expressed in terms of a Lense-Thirring precession velocity $\omega_{LT_{CG}}$, by
\begin{equation}
\delta\tau_0=8\frac{\omega_{LT_{CG}}}{c^2}\frac{\pi R^2}{\sqrt{p + \frac{u}{R} + vR - kR^2}},
\end{equation}
where
\begin{equation}
\omega_{LT_{CG}}=jc\left(\frac{u+\left(p-1\right)R+vR^2-kR^3}{2R^3}\right).
\label{LT_result}
\end{equation}
Using the values for the constants given in Eq.(\ref{gr_limit}), this gives the general relativistic result obtained in Ref. \cite{Tartaglia}.



\section{V. Discussion and conclusion}
In this paper, we first considered how circular orbits behave in conformal gravity, thereby obtaining a generalization of Kepler's third law.  The main results of this paper are obtained in Eq.(\ref{precession}) and Eq.(\ref{LT_result}), which represent the geodetic and Lense-Thirring effects in conformal gravity respectively.
\newline
Apart from the Einstein term $3\pi\beta/R$, Eq.(\ref{precession}) also includes terms due to the linear term $\gamma r$ in the Mannheim-Kazanas metric as
well as the cosmological term $kr^2$. We note that as in the case of the other already studied classical tests, namely the bending of light \cite{sultana10},
time delay \cite{edery98} and perihelion precession \cite{sultana12}, the contribution to the geodetic precession from the linear term in the metric has an
opposite sign to that of the Einstein term, so that it reduces the amount of precession per orbit. Moreover this contribution increases linearly with the radius $R$ of the circular orbit, such that at larger radii its effect can cancel that of the Einstein term.
\newline
The Lense-Thirring precession velocity was obtained in terms of the Sagnac time delay for a stationary observer which occurs solely due to the drag of spacetime by the rotating central object. Assuming that the constants $u$, $p$ and $k$ in (\ref{LT_result}) take the same values as in the Kerr-de Sitter solution, we see that the conformal gravity correction through the constant $v$ [which corresponds to $\gamma$ when $j=0,\ \theta = \pi/2$ in (\ref{kerr-like})]  has again a diminishing effect on the total precession velocity. However in this case this correction is inversely proportional to the radius of the circular orbit.
\newline

For a gyroscope in the Earth's orbit, such as GP-B, the extra contribution to the geodetic precession from the linear and cosmological terms in the metric are insignificant for practical purposes, when the value of $\gamma$ is taken as the reciprocal of the Hubble radius, as required for the fitting of galactic rotational curves shown in Ref. \cite{conformal1}.
In earlier studies such as those in Refs.\cite{sultana12,mannheim11} constraints were obtained for the conformal gravity parameter $\gamma$. However in both
cases of geodetic and Lense-Thirring effects the uncertainties in available GP-B data are too large to obtain reasonable
constraints. So, for example, using the uncertainty in the observed value for the GP-B geodetic drift mentioned in the introduction, and the Weyl's
gravity correction to geodetic effect in (\ref{precession}), one gets $\gamma \leq 1.5\times10^{-20}\ \mathrm{cm}^{-1}$, which is eight orders of magnitude
larger than the value obtained from galactic rotational curves.
\newline

It should be remarked, as we did in the introduction, that the nature (and therefore the magnitude) of the constant $\gamma$ is still unknown. So when $\beta = 0$ or when $r$ is sufficiently large so that all $\beta$ dependent terms in (\ref{eq:metric}) can be ignored, the metric can be rewritten in the form
\begin{eqnarray}
ds^2 & = & \textstyle\frac{[1 - \rho^2(\gamma^2/16 +
k/4)]^2}{R^2(\tau)[(1 - \gamma\rho/4)^2 + k\rho^2/4]^2}
[-d\tau^2\nonumber \\
& + & \textstyle\frac{R^2(\tau)}{[1 - \rho^2(\gamma^2/16 +
k/4)]^2}\textstyle(d\rho^2 + \rho^2\,d\Omega_{2}^{2})],\nonumber \\
\end{eqnarray}
where
\begin{equation}
\rho = \frac{4r}{2(1 + \gamma r - k r^2)^{1/2} + 2 + \gamma r},
\end{equation}
and
\begin{equation}
\tau = \int R(t)dt.
\end{equation}
This is conformally related to the FLRW metric with arbitrary scale factor $R(\tau)$ and spatial curvature $\kappa = - k - \gamma^2/4$. Therefore, one may interpret the Mannheim-Kazanas metric as describing a spherically symmetric object embedded in a conformally flat background space. Now the fact the curvature of this background space depends on $\gamma$ and $k$, points toward a cosmological origin of $\gamma$, which in our case is substantiated by the fact that its contribution to the geodetic precession in (\ref{precession}) is independent from the mass of the central object, just like the cosmological term. However having said this, there is nothing in the theory that forbids $\gamma$ from being also system dependent, in which case it may provide the necessary changes in the spacetime geometry to allow the embedding of a spherically symmetric matter distribution in a cosmological background.

To conclude, this paper gives a derivation of Kepler's third law and the gravitomagnetic effects in the alternative gravitational theory of conformal gravity. On the solar system scale, where conformal gravity (like other alternative gravitational theories) agrees with general relativity, the corrections to the Einstein precession are very small to be measured by any space-based experiment. However these may become significant on much larger scales, especially when the geodetic effect is considered.

\section{Acknowledgments}
J. L. S. wishes to thank the Physics Department at the University of Malta for hospitality during the completion of this work.

\end{document}